\RequirePackage{lineno}
\documentclass[aps,prc,twocolumn,showpacs,superscriptaddress,groupedaddress,floatfix]{revtex4}  
\usepackage{graphicx}  
\usepackage{amssymb}
\usepackage{amsmath}
\usepackage{subfigure}
\usepackage{multirow}
\numberwithin{table}{section}

\begin{document}


\title{Wide acceptance measurement of the K$^- \slash \text{K}^+$ ratio from Ni+Ni collisions at 1.91A GeV}

\def\wars{Institute of Experimental Physics, Faculty of Physics, University of Warsaw, Warsaw, Poland}
\def\heid{Physikalisches Institut der Universit\"{a}t Heidelberg, Heidelberg, Germany}
\def\darm{GSI Helmholtzzentrum f\"{u}r Schwerionenforschung GmbH, Darmstadt, Germany}
\def\seou{Korea University, Seoul, Korea}
\def\cler{Laboratoire de Physique Corpusculaire, IN2P3/CNRS, and Universit\'{e} Blaise Pascal, Clermont-Ferrand, France}
\def\zagr{Ru{d\llap{\raise 1.22ex\hbox{\vrule height 0.09ex width 0.2em}}\rlap{\raise 1.22ex\hbox{\vrule height 0.09ex width 0.06em}}}er Bo\v{s}kovi\'{c} Institute, Zagreb, Croatia}
\def\munI{Excellence Cluster Universe, Technische Universit\"{a}t M\"{u}nchen, Garching, Germany}
\def\munII{E62, Physik Department, Technische Universit\"{a}t M\"{u}nchen, Garching, Germany}
\def\vien{Stefan-Meyer-Institut f\"{u}r subatomare Physik, \"{O}sterreichische Akademie der Wissenschaften, Wien, Austria}
\def\sp{University of Split, Faculty of Science, Split, Croatia}
\def\buda{Wigner RCP, RMKI, Budapest, Hungary}
\def\mosc{Institute for Theoretical and Experimental Physics, Moscow, Russia}
\def\dres{Institut f\"{u}r Strahlenphysik, Helmholtz-Zentrum Dresden-Rossendorf, Dresden, Germany} 
\def\harb{Harbin Institute of Technology, Harbin, China}
\def\kurc{National Research Centre "Kurchatov Institute", Moscow, Russia}
\def\buch{Institute for Nuclear Physics and Engineering, Bucharest, Romania}
\def\stra{Institut Pluridisciplinaire Hubert Curien and Universit\'{e} de Strasbourg, Strasbourg, France}
\def\tsing{Department of Physics, Tsinghua University, Beijing 100084, China}
\def\lan{Institute of Modern Physics, Chinese Academy of Sciences, Lanzhou, China}

\author{K.~Piasecki}\email{krzysztof.piasecki@fuw.edu.pl} \affiliation{\wars}
\author{N.~Herrmann} \affiliation{\heid}
\author{R.~Averbeck} \affiliation{\darm}
\author{A.~Andronic} \affiliation{\darm} 
\author{V.~Barret} \affiliation{\cler} 
\author{Z.~Basrak} \affiliation{\zagr} 
\author{N.~Bastid} \affiliation{\cler}
\author{M.L.~Benabderrahmane} \affiliation{\heid}
\author{M.~Berger} \affiliation{\munI} \affiliation{\munII}
\author{P.~Buehler} \affiliation{\vien} 
\author{M.~Cargnelli} \affiliation{\vien} 
\author{R.~\v{C}aplar} \affiliation{\zagr}
\author{E.~Cordier} \affiliation{\heid}
\author{P.~Crochet} \affiliation{\cler} 
\author{O.~Czerwiakowa} \thanks{Present address: National Centre for Nuclear Research, Otwock, Poland} \affiliation{\wars} 
\author{I.~Deppner} \affiliation{\heid}
\author{P.~Dupieux} \affiliation{\cler}
\author{M.~D\v{z}elalija} \affiliation{\sp}
\author{L.~Fabbietti} \affiliation{\munI} \affiliation{\munII}
\author{Z.~Fodor} \affiliation{\buda}
\author{P.~Gasik} \affiliation{\wars} \affiliation{\munI} \affiliation{\munII}
\author{I.~Ga\v{s}pari\'c} \affiliation{\zagr}
\author{Y.~Grishkin} \affiliation{\mosc}
\author{O.N.~Hartmann} \affiliation{\darm}
\author{K.D.~Hildenbrand} \affiliation{\darm}
\author{B.~Hong} \affiliation{\seou}
\author{T.I.~Kang} \affiliation{\seou}
\author{J.~Kecskemeti} \affiliation{\buda}
\author{Y.J.~Kim} \affiliation{\darm}
\author{M.~Kirejczyk} \thanks{Present address: National Centre for Nuclear Research, Otwock, Poland} \affiliation{\wars} 
\author{M.~Ki\v{s}} \affiliation{\darm} \affiliation{\zagr}
\author{P.~Koczon} \affiliation{\darm}
\author{M.~Korolija} \affiliation{\zagr}
\author{R.~Kotte} \affiliation{\dres}
\author{A.~Lebedev} \affiliation{\mosc}
\author{Y.~Leifels} \affiliation{\darm}
\author{A.~Le F\`{e}vre} \affiliation{\darm}
\author{J.L.~Liu} \affiliation{\heid} \affiliation{\harb}
\author{X.~Lopez} \affiliation{\cler}
\author{A.~Mangiarotti} \affiliation{\heid}
\author{V.~Manko} \affiliation{\kurc}
\author{J.~Marton} \affiliation{\vien}
\author{T.~Matulewicz} \affiliation{\wars}
\author{M.~Merschmeyer} \affiliation{\heid}
\author{R.~M\"{u}nzer} \affiliation{\munI} \affiliation{\munII}
\author{D.~Pelte} \affiliation{\heid}
\author{M.~Petrovici} \affiliation{\buch}
\author{F.~Rami} \affiliation{\stra}
\author{A.~Reischl} \affiliation{\heid}
\author{W.~Reisdorf} \affiliation{\darm}
\author{M.S.~Ryu} \affiliation{\seou}
\author{P.~Schmidt} \affiliation{\vien}
\author{A.~Sch\"{u}ttauf} \affiliation{\darm}
\author{Z.~Seres} \affiliation{\buda}
\author{B.~Sikora} \affiliation{\wars}
\author{K.S.~Sim} \affiliation{\seou}
\author{V.~Simion} \affiliation{\buch}
\author{K.~Siwek-Wilczy\'{n}ska} \affiliation{\wars}
\author{V.~Smolyankin} \affiliation{\mosc}
\author{G.~Stoicea} \affiliation{\buch}
\author{K.~Suzuki} \affiliation{\vien}
\author{Z.~Tymi\'{n}ski} \thanks{Present address: National Centre for Nuclear Research, Otwock, Poland} \affiliation{\wars} 
\author{P.~Wagner} \affiliation{\stra}
\author{I.~Weber} \affiliation{\sp} 
\author{E.~Widmann} \affiliation{\vien}
\author{K.~Wi\'{s}niewski} \affiliation{\wars} 
\author{Z.G.~Xiao} \affiliation{\tsing}
\author{H.S.~Xu} \affiliation{\lan}
\author{I.~Yushmanov} \affiliation{\kurc}
\author{Y.~Zhang} \affiliation{\lan}
\author{A.~Zhilin} \affiliation{\mosc}
\author{V.~Zinyuk} \affiliation{\heid}
\author{J.~Zmeskal} \affiliation{\vien}

\collaboration{FOPI Collaboration} \noaffiliation

\date{\today}

\begin{abstract}

The FOPI Collaboration at the GSI SIS-18 synchrotron measured charged kaons
from central and semi-central collisions of Ni+Ni at a beam energy of 1.91A GeV. 
We present the distribution of the K$^- \slash \text{K}^+$ ratio on the 
energy vs polar angle plane in the nucleon-nucleon center-of-mass frame,
with and without subtraction of the contribution 
of $\phi$(1020) meson decays to the K$^-$ yield.
The acceptance of the current experiment is substantially wider
compared to the previous measurement of the same colliding system.
The ratio of K$^-$ to K$^+$ energy spectra is expected to be sensitive to 
the in-medium modifications of basic kaon properties like mass.
Recent results obtained by the HADES Collaboration at 1.23A and 1.76A GeV indicate
that after inclusion of the $\phi$ meson decay contribution to the K$^-$ production
no difference between the slopes of the K$^-$ and K$^+$ energy spectra is observed
within uncertainties.
For our data a linear fit to this ratio obtained after subtraction
of the $\phi$ meson contribution still shows a decrease with kinetic energy,
although a constant value cannot be rejected.
The contribution of $\Lambda(1520) \rightarrow p\text{K}^-$ decays 
estimated from fitting the thermal model to the experimental yields 
appears to be another factor of moderate relevance.

\end{abstract}

\pacs{25.75.Dw, 13.60.Le}
\maketitle


\section{Introduction}
\label{Sect:intro}

Modifications of the basic properties of kaons (like mass or decay constant)
inside a hot and dense nuclear medium have been the subject of intensive study 
and debate throughout the last 30 years
~\cite{Kapl86,Brow91,Weis96,Scha97,Cass97,Ramo00,Tolo02,Lutz04,Fuch06,Hart12}.
It is predicted that under such conditions the system should tend toward
the partial restoration of chiral symmetry.

In the earlier approaches these modifications were parameterized 
in terms of the kaon-nucleus potential $U_{\text{KN}}$ 
and the effective mass $m^*$~\cite{Weis96,Scha97,Cass97,Fuch06,Hart12}. 
While there are no universal definitions of these quantities, 
in this paper we follow a practical approach, where
$m^*$ is the energy of a particle at rest in nuclear matter at normal density and 
$U_{\text{KN}}$ is the difference between $m^*$ and mass in vacuum ($m$).
Calculations predicted that the potential 
for a kaon should be positive and that for an antikaon negative. 
Within the latter approaches it is still predicted that kaons (K$^+$ and K$^0$)
remain ``good'' quasiparticles with narrow width.
However, the antikaons (K$^-$ and $\bar{K^0}$) exhibit a non-trivial structure 
of self-energy, and the {\it potential} should be only approximately
understood as the average of this structure~\cite{Ramo00,Tolo02,Lutz04}.
In particular, the most probable K$^-$ production channel around threshold
is $\pi\text{Y} \rightarrow \text{K}^-$N, 
where Y denotes the hyperon ($\Lambda$ or $\Sigma$).
This channel is predicted to have an intermediate step in medium involving 
$\Lambda^*(1405)$ or $\Sigma^*(1385)$, where the production of the latter particle
under these conditions was experimentally confirmed~\cite{Lope07}.

If nuclei are collided at beam energies around the thresholds 
for the production of the respective K mesons in a nucleon-nucleon (NN) collision
(about 1.6 GeV for K$^+$ and 2.5 GeV for K$^-$), 
the probability of producing a kaon per event is around 1\%~\cite{Fors07}.
Thus, events with open-strangeness production usually contain only one kaon, 
which can be used as a probe of the dynamics in the nuclear medium.
These collisions have been intensively studied in particular 
at the SIS-18 accelerator, delivering heavy-ion beams up to an energy of 2A GeV.

When the kaon leaves the hot and condensed collision zone, where it was produced, 
its effective mass must return to the vacuum value.
As the simplest energy reservoir is the kinetic energy,
it is predicted that K$^+$ leaving the centers of density should accelerate, 
and K$^-$ should decelerate.
Thus, the ratio of K$^- \slash \text{K}^+$ 
as a function of kinetic energy is expected to become steeper
if the in-medium effects occur as predicted.
Also, the attraction toward centers of density should cause K$^-$ 
to exhibit a flow pattern similar to that of protons, 
whereas for K$^+$ the effect is predicted to be opposite~\cite{Liko95}.

However, in course of the propagation of kaons in the heavy-ion collision zone,
the effects of modifications of basic kaon properties compete with other 
phenomena like absorption (affecting mainly K$^-$ via 
K$^- \text{N} \rightarrow \pi$Y) or rescattering from surrounding nucleons.
In addition, the K$^-$ spectra are fed by the dominant $\phi$(1020) meson 
decay channel, $\phi \rightarrow \text{K}^+\text{K}^-$ (BR~=~48.9\%~\cite{PDG}),
as the threshold for $\phi$ production is very close to that for the K$^+$K$^-$ pair.	

A search for in-medium effects in the flow pattern
was recently reported~\cite{Ziny14}. A comparison of
the rapidity dependence of the $v_1$ coefficient of charged kaons emitted
from Ni+Ni collisions at 1.9A GeV to the predictions of the
IQMD and HSD transport models~\cite{IQMD,HSD} pointed to rather small values of
$U_{\text{K}^+\text{N}}$, between 0 and 20 MeV. For K$^-$ the IQMD prediction with
$U_{\text{K}^-\text{N}} = -45$~MeV reproduced the $v_1$ pattern, 
whereas the HSD calculations employing the \mbox{G-Matrix} formalism 
corresponding to $U_{\text{K}^-\text{N}} = -50$~MeV 
overestimated the experimental values.
Concerning the comparison of the transverse momentum dependence of $v_1$
for K$^+$, the transport model predictions again pointed to values 
between 0 and 20 MeV, although none of these models reproduced 
the trend of the data points in full.

Studies of in-medium effects via the experimentally measured 
K$^- \slash \text{K}^+$ ratio were presented in~\cite{Laue00,Wisn00,Gasi16}. 
Whereas these analyses demonstrated the sensitivity 
of this observable to in-medium effects, 
they were hampered either by low statistics, 
lack of (or insufficient) inclusion of the $\phi$(1020) meson decay feeding, 
or very narrow acceptance.
A comparison of four data points for the energy dependence of the 
K$^- \slash \text{K}^+$ ratio from Al+Al collisions at 1.9A GeV
to the HSD transport model calculations with 
\mbox{U$_{\text{K}^+\text{N}} = 40$~MeV} and 
\mbox{U$_{\text{K}^-\text{N}} = -50$~MeV}~\cite{Gasi16} 
initially appeared to reproduce the data successfully. 
However, they did not account for the feeding from $\phi$ mesons.
The available samples of $\phi$ in its dominant decay channel,
$\phi \rightarrow \text{K}^+\text{K}^-$ (BR~=~48.9\%~\cite{PDG}),
were quite scarce (100-170 events~\cite{Agak09,Pias15,Gasi16,Pias16}).
Despite this, the contribution of their decays to the K$^-$ spectra
was found significantly to reduce the slope of the energy spectrum~\cite{Lore10,Pias15,Gasi16,Adam18}
and therefore compete with the effect of the attractive $\bar{K}$N potential. 
After correcting for the $\phi$ meson contribution to the K$^-$ spectra,
the HSD prediction with non-vanishing KN potentials seemed to overestimate the data.

On the other hand, one may put forward the hypothesis, 
that the difference between the slopes of the energy distributions 
of K$^-$ and K$^+$ mesons can be fully explained 
by feeding of $\phi$ meson decays to the K$^-$ meson distribution.
This approach was considered by the HADES Collaboration for 
Ar+KCl at 1.76A GeV~\cite{Lore10} and for Au+Au at 1.23A GeV~\cite{Adam18},
and was found to be consistent with the experimental data within errors.

The distributions of the kaon polar angle, the second phase space dimension,
were often found to deviate from isotropy~\cite{Fors07,Gasi16}. 
For most of the investigated systems the $a_2$ anisotropy coefficients
for K$^+$ and K$^-$ were found to be equal within 3 standard deviations.
However, globally the values of $a_2$ for K$^+$ appear to be 
somewhat larger than those for K$^-$. 
Thus, in the phase space distribution of the K$^- \slash \text{K}^+$ ratio,
the anisotropy effects for K$^+$ and K$^-$ may not cancel out. 
Additionally, analysis of the polar distribution within the 
IQMD transport model showed that the degree of anisotropy should be 
sensitive to both the potential and rescattering effects~\cite{Hart12}. 
Thus, a measurement of this ratio not only as a function of energy
but also of polar angle could deliver a more precise probe of these phenomena. 
It should also be supplemented by a measurement of the 
contribution of $\phi$(1020) mesons to the K$^-$ spectrum obtained for the same reaction.

The FOPI collaboration has addressed this goal, in particular benefitting
from the \mbox{MMRPC} detector, a Time-of-Flight device
characterized by high granularity and excellent timing properties~\cite{Kis11}.
In this paper we present the kinematic distributions and production ratio
of charged kaons emitted from central and semi-central Ni+Ni collisions 
at a beam kinetic energy of 1.91A GeV (the same experiment as for the
K$^\pm$ flow study~\cite{Ziny14} and $\phi$ meson analysis in~\cite{Pias15}).
The advantages of this data sample with respect to earlier analyses of the
K$^- \slash \text{K}^+$ ratio are the considerably wider acceptance
and the additional experimental information on the $\phi$ meson production.

\section{Experiment}
\label{Sect:exp}

A detailed description of the FOPI spectrometer was given in~\cite{FOPI}, 
and the experiment was reported in~\cite{Ziny14,Pias15}.
Here we highlight only those features most relevant to the present analysis.

The innermost detector of the FOPI apparatus is the Central Drift Chamber (CDC)
covering the polar angles ($27^\circ < \vartheta_{\text{lab}} < 113^\circ$)
\footnote{The angles are given with respect to the target position}. 
It is encircled by two time-of-flight (ToF) devices, 
the Plastic Scintillation Barrel (PSB), spanning
$55^\circ < \vartheta_{\text{lab}} < 110^\circ$, 
and the Multi-strip Multi-gap Resistive Plate Counter~\cite{Kis11} (MMRPC), 
mounted at $30^\circ < \vartheta_{\text{lab}} < 53^\circ$.
This setup is surrounded by a magnet generating a solenoidal
field of B = 0.617~T. 
The forward polar angles, $6.5^\circ < \vartheta_{\text{lab}} < 23^\circ$,
are covered by the Plastic Wall (PlaWa) detector. 

A $^{58}$Ni beam with a kinetic energy of 1.91A GeV was incident 
on a $^{58}$Ni target of 1\% interaction probability.  
The medium-bias trigger required the minimum multiplicity of
charged hits in the PlaWa (PSB) to be $\geq$~5 ($\geq$~1). 
This allowed us to select the sample of central and semi-central collisions 
corresponding to $(56 \pm 3)~\% $ of the total geometrical cross section.
Within these conditions $7.6\times 10^7$ events were collected.
Assuming the sharp cut-off approach and the geometrical model, 
the average number of participant nucleons was estimated to be 
$\langle A_{\text{part}} \rangle_{\text{b}} = 46.5 \pm 2.0$
(c.f. Appendix A of~\cite{Pias16}).

\section{Phase Space Distributions}
\label{Sect:ana}

\subsection{Raw kaon spectra}
The particle identification methods used in our experiment were described 
in Refs.~\cite{Ziny14,Pias15}. Here only the key points are highlighted.
For each event the {\it tracks} of particles traversing the CDC are reconstructed 
from the activated wires ({\it ``hits''}). A reconstruction of the
vertex position allows us to reject reactions occurring outside the target.
In the next step, ``good track candidates'' are selected
by requiring a minimum multiplicity of hits in a track
and a maximum distance between the track and the vertex.
Fitting a helical curve to the series of hits marked by a particle 
in a solenoidal magnetic field allows us to obtain the momentum vector $\vec{p}$. 
The amplitude of the signals from activated wires is used to measure 
the specific energy loss. 
Correlating these two observables allows us to 
identify many of the charged emission products
and extract the CDC-based mass parameter, $m_{\text{CDC}}$.
This procedure is, however, insufficient for most of the investigated K$^\pm$ mesons. 
Additional information is obtained from either of the two installed
ToF devices: PSB and MMRPC. A combination of time of flight and
path length allows us to obtain the velocity $v$ of the particle emitted from
the target. A histogram of identified tracks on the $p - v$ plane is shown 
in Fig.~1 of Ref.~\cite{Pias15}. Substituting $p$ and $v$ into the
relativistic dependency $p = m\gamma v$, where $\gamma$ is the Lorentz factor, 
allows us to extract the particle mass parameter, $m_{\text{ToF}}$.
The distributions of $m_{\text{ToF}}$, shown for the same data in Fig.~1c,d 
of Ref.~\cite{Ziny14}, clearly exhibit a peak around the nominal mass of charged kaon.

\begin{figure}[tb]
 \includegraphics[width=\linewidth]{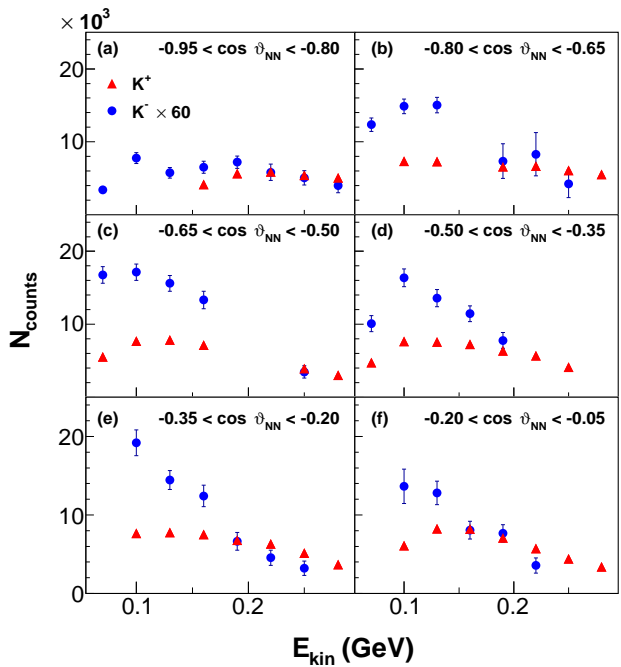}
 \caption{\label{fig:EkinRaw}(Color online)
   Raw kinetic energy spectra of charged kaons emitted from central and semi-central 
   collisions of Ni+Ni at a beam energy of 1.91A GeV
   within six bins of $\cos \vartheta_{\text{NN}}$.
   Red triangles correspond to K$^+$ and blue full circles to K$^-$.
   The yield of K$^-$ was multiplied by 60.
 }
\end{figure}

To minimize edge effects, the range of accepted polar angles 
was trimmed down to $30^\circ < \vartheta_{\text{lab}} < 53^\circ$.
In addition, to account for the limits of detection of low-p$_{\text{T}}$ particles
by the CDC-ToF pair of detectors and some slight inconsistency 
in the reproduction of the detection capability within the 
GEANT3~\cite{GEANT} environment in this region, 
the K$^+$ (K$^-$) candidates were required to have a transverse
momentum $p_{\text{T}}$ of at least 0.18 (0.14) GeV/$c$. 

The raw spectrum of measured kaons, shown in Fig.~\ref{fig:EkinRaw},
was obtained with 232300 measured K$^+$ and 5660 K$^-$ mesons.
A common multiplicative factor of 60 was applied to the K$^-$ data points
in order to present both profiles on one plot.

\subsection{Efficiency determination}
\label{Sect:eff}

The efficiency correction for charged kaons on the 
$E^{\text{kin}}_{\text{NN}} - \cos \vartheta_{\text{NN}}$ plane 
was obtained in two stages:
via GEANT3-based simulations and an additional procedure intended
to extract and apply the internal efficiency of the ToF detectors.
 
\begin{figure}[tb]
 \includegraphics[width=\linewidth]{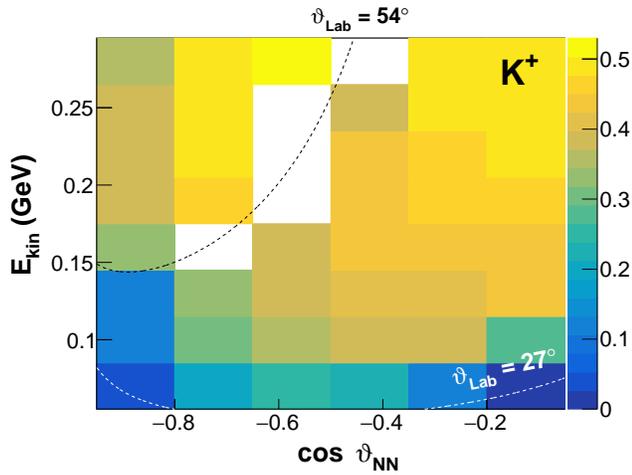}
 \caption{\label{fig:Eff}(Color online) 
   GEANT3-based part of the efficiency map 
   on the $E^{\text{kin}}_{\text{NN}} - \cos \vartheta_{\text{NN}}$ plane
   for detection of charged kaons from Ni+Ni collisions at 1.91A GeV within FOPI. 
   Dotted curves mark the geometric boundaries of the MMRPC 
   and Plastic Barrel detectors. See text for details.
 }
\end{figure}

In the first stage kaons were sampled from the homogeneous distribution
on this plane. They were subsequently added to the events 
of Ni+Ni collisions generated within the IQMD transport code~\cite{IQMD},
which aims to reproduce the realistic background of particles emitted
from the heavy-ion collisions. Within the GEANT3 environment
particles were transported to the detection modules of the virtual FOPI setup.
The hadronic interactions of kaons with the traversed medium were switched on.
After the detector responses were simulated, the events were processed
by the same tracking and matching routines as for the true experimental data.
The resulting efficiency maps are shown in Fig.~\ref{fig:Eff}. 
The elongated drop of efficiency toward lower kinetic energy and 
$\cos \vartheta_{\text{NN}}$ is due to the decay of kaons in flight.

However, the digitization routines of the GEANT3 package did not account for
the intrinsic efficiency of the ToF devices. A determination of this efficiency
was made by a dedicated procedure, described in Sect. IV B of Ref.~\cite{Pias15},
where the maps are shown in the lower panels of Fig.~3.
This efficiency component was included by weighting the experimental kaon
signal with the appropriate factors.

\begin{figure}[tb]
 \includegraphics[width=\linewidth]{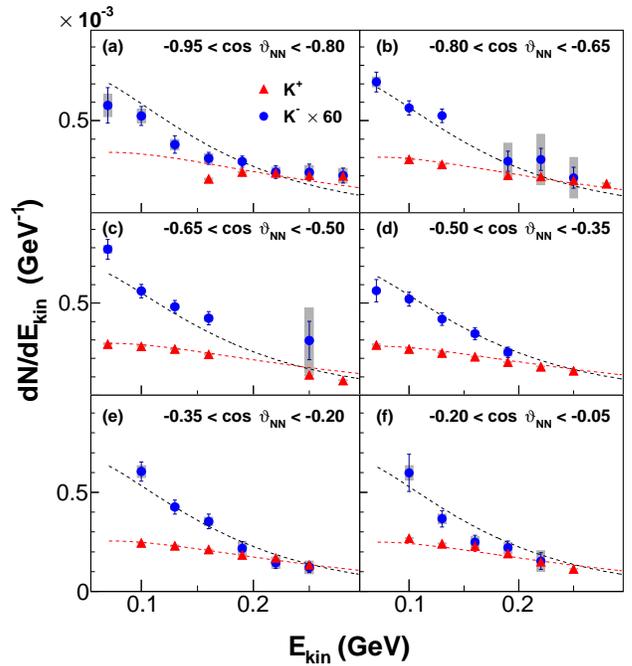}
 \caption{\label{fig:EkinReco}(Color online)
   Kinetic energy spectra of charged kaons emitted from
   central and semi-central collisions of Ni+Ni at a beam energy of 1.91A GeV
   for subsequent bins of $\cos \vartheta_{\text{NN}}$.
   Red triangles correspond to K$^+$ and blue full circles to K$^-$.
   The yield of K$^-$ was multiplied by 60.
   Grey boxes denote systematic errors.
   The best fits of Eq.~\ref{EQ:fitBoltzAni} to the data are shown as dashed lines.
 }
\end{figure}

\section{Results}
\label{Sect:resu}

\begin{figure*}
 \begin{minipage}[c]{0.495\textwidth}
  \begin{center}  
   \includegraphics[width=0.97\textwidth]{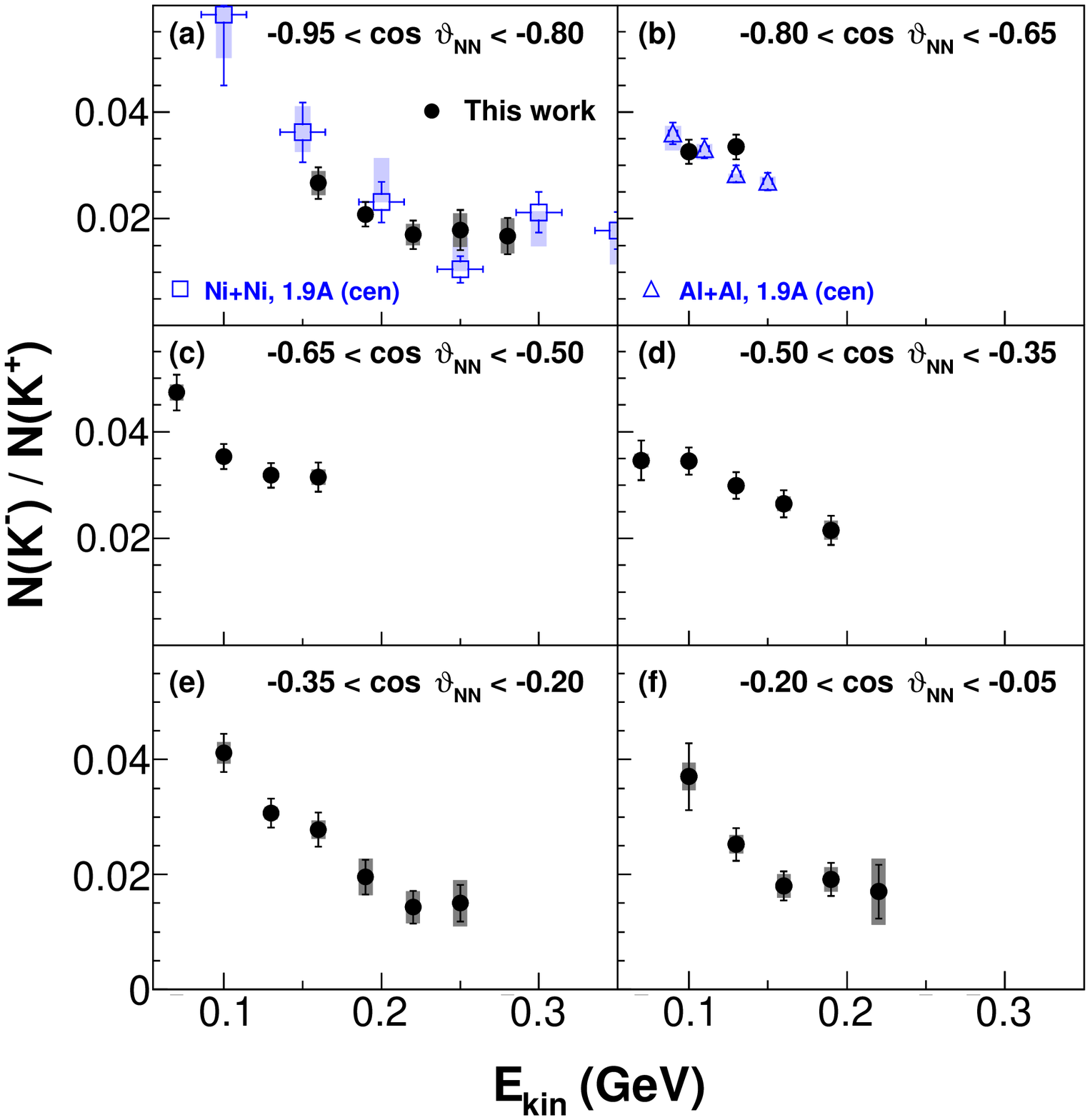}
  \end{center}  
 \end{minipage}
 \begin{minipage}[c]{0.495\textwidth}
  \begin{center}  
   \includegraphics[width=0.97\textwidth]{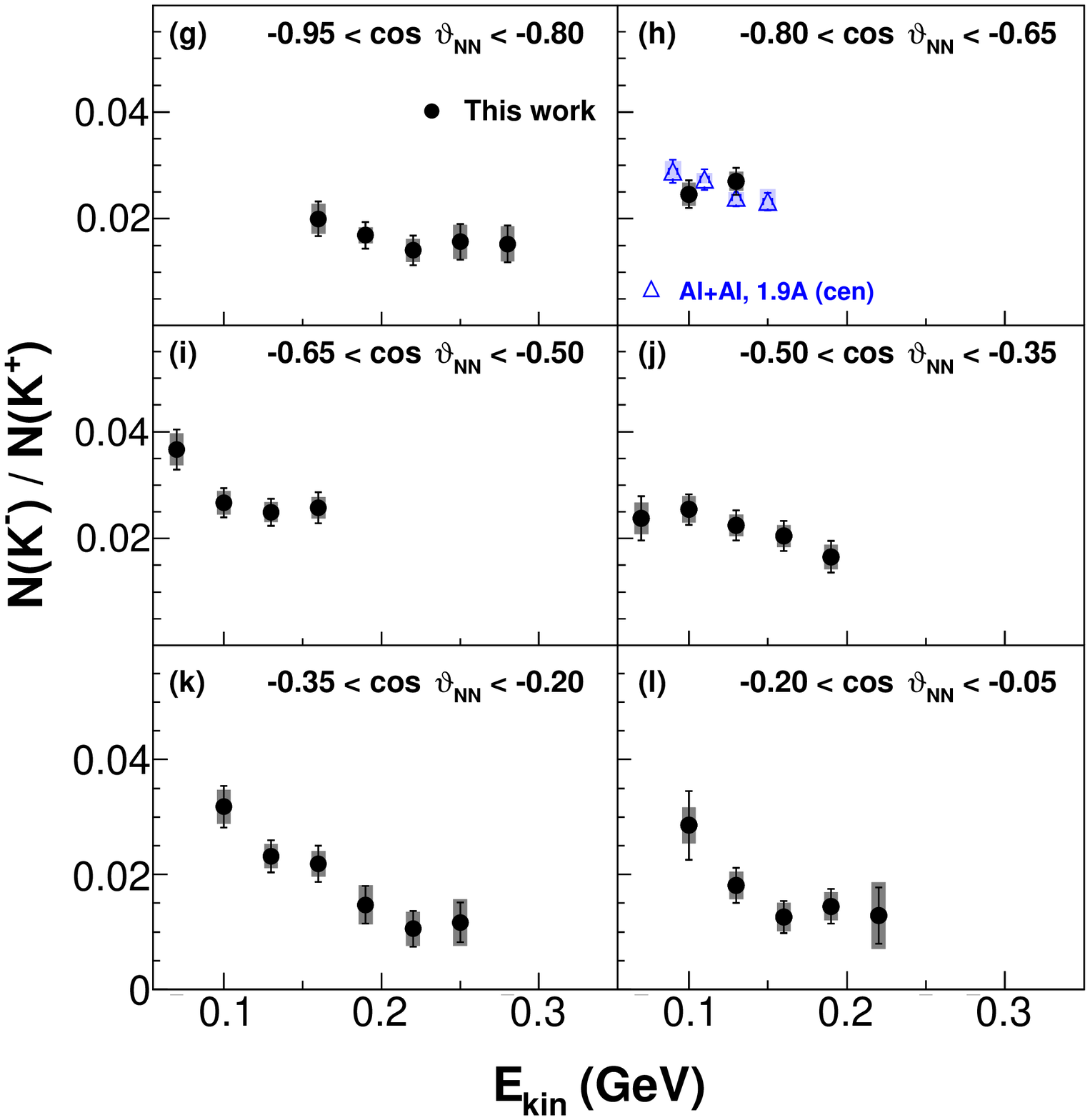}
  \end{center}  
 \end{minipage}
 \caption{\label{fig:kmkpratio}(Color online) 
   Black circles: kinetic energy distribution of the K$^- \slash \text{K}^+$ ratio 
   for kaons emitted from central and semi-central collisions of Ni+Ni at 1.91A GeV,
   within six bins of $\cos \vartheta_{\text{NN}}$.
   Grey boxes denote systematic errors.
   For the data in left-hand panels the contribution of the $\phi$ decays 
   to K$^-$ production is included.
   The right-hand panels correspond to the same distribution after subtraction
   of this contribution from the K$^-$ spectra.
   Open squares: data for central Ni+Ni collisions from the previous experiment~\cite{Wisn00}.
   Open triangles: data for central Al+Al collisions~\cite{Gasi16}.
   See text for details.
 }
\end{figure*}

The distributions of charged kaons on the 
$E^{\text{kin}}_{\text{NN}}$ - $\cos \vartheta_{\text{NN}}$ plane 
are shown in Fig.~\ref{fig:EkinReco} 
(note the factor of 60 multiplying the K$^-$, as mentioned in Sect.~\ref{Sect:ana}).
The same binning was used as for Fig.~\ref{fig:EkinRaw}.
The K$^-$ kinetic energy spectra appear to be softer than those of K$^+$.
This finding is very common for charged kaons emitted from heavy-ion
collisions at similar energies~\cite{Fors07}.
The distribution of the K$^- \slash \text{K}^+$ ratio obtained
is presented in Fig.~\ref{fig:kmkpratio}. 

The leading contributions to the systematic errors were found to be: 
\begin{itemize}
\item sensitivity to the selection of the minimum number of CDC hits forming a ``good track'', 
\item choice of the background function (linear or exponential) 
  under the kaon peak in the mass spectrum,
\item minimum cutoff value of the $m_{\text{CDC}}$ parameter 
  in the case of tracks matched with Plastic Barrel hits,
\item binning of the spectrum of $m_{\text{ToF}}$.
\end{itemize}
By varying these conditions, slightly different values for the distributions
of the K$^+$ and K$^-$ yields and the \mbox{K$^-\slash \text{K}^+$} ratio 
were obtained. In this way for each i-th point on the 
$E^{\text{kin}}_{\text{NN}} - \cos \vartheta_{\text{NN}}$ plane
a distribution $P_\text{i}$ of values of an investigated quantity was generated. 
The final result for each i-th point was determined by averaging the values of $P_\text{i}$. 
This approach also allowed us to select the confidence level (CL) at which
the systematic errors were estimated. In this analysis we chose CL = 68.3\% 
(corresponding to 1$\sigma$ of the Gaussian distribution), 
based directly on the distributions $P_\text{i}$.
It should be noted, that this procedure was applied independently
for the kaon yields and the K$^-\slash \text{K}^+$ ratio.

Previously, two kinetic energy distributions of the
\mbox{K$^- \slash \text{K}^+$} ratio
were measured at 1.9A GeV within narrow $\cos \vartheta_{\text{NN}}$ ranges. 
The distribution for the previous experiment with Ni+Ni 
with a slightly different FOPI setup and ToF detectors was measured within
$-0.97 < \cos \vartheta_{\text{NN}} < -0.87$~\cite{Wisn00},
and is shown in panel (a) of Fig.~\ref{fig:kmkpratio}.
The data from Al+Al collisions, obtained within
$-0.87 < \cos \vartheta_{\text{NN}} < -0.72$~\cite{Gasi16},
are also plotted placed in panel (b).
We find our data consistent with the previously obtained results,
but covering a much broader acceptance. 
The complete data set for the phase space distributions of the charged kaons 
and the K$^- \slash \text{K}^+$ ratio is listed in
Table~\ref{Tab:KaonYields} in Appendix~\ref{Sect:Data}.

\section{Discussion}

\subsection{Phase space analysis of charged kaons}

The efficiency-corrected phase space distribution of K$^\pm$ mesons, 
reported in Sect.~\ref{Sect:resu} was fitted with the following \textit{ansatz},

\begin{equation}
 \begin{aligned}
 \label{EQ:fitBoltzAni}
 \frac{d^2N}{dE^{\text{kin}}_{\text{NN}}~d\cos \vartheta_{\text{NN}}}
 ~=~ & \frac{N}{C}
       \cdot p_{\text{NN}} ~E_{\text{NN}} ~\exp \left( -E_{\text{NN}} / T_{\text{eff}} \right) \\
     & \cdot \left( 1 + a_2 \cos^2 \vartheta_{\text{NN}} \right)
 \end{aligned}
\end{equation}

\noindent which consists of the Boltzmann-like kinetic term parameterized by the
effective temperature $T_{\text{eff}}$, multiplied by the angular anisotropy term,
where $a_2$ is the polar anisotropy coefficient. $N$ is the yield of emitted kaons
per triggered collision, and $C$ is the normalization constant, defined so that
the integral of Eq.~\ref{EQ:fitBoltzAni} yields $N$.
Since the energy and polar angle are not correlated in this formula, 
the fit to the data shown in Fig.~\ref{fig:EkinReco} also allows us to extract 
$N$ directly, with an uncertainty $\Delta N$ free from correlation terms. 
The results of this procedure, shown in Table~\ref{Tab:KaonFit}, confirm that
the inverse slope of the energy spectrum of K$^-$ is smaller than that of K$^+$.
Despite the fact that the parameterization of the distribution by Eq.~\ref{EQ:fitBoltzAni} 
is not the same as that of Eq.~3 in Ref.~\cite{Pias15} applied to the same K$^-$ data, 
the slope obtained here is in agreement with the profile
of slopes shown in Fig.~6 of that paper. It also agrees within 2$\sigma$ with the slope
obtained by the KaoS Collaboration for the inclusive reactions~\cite{Fors07}.
Concerning the $a_2$ coefficients, they appear to be somewhat smaller than the values
for inclusive Ni+Ni collision, obtained by KaoS (c.f. Table II of Ref.~\cite{Fors07}),
however, the minimum bias triggers were not defined identically.
It also has to be noted that the $\chi^2 \slash \nu$ value for our fit to the K$^+$ 
distribution is considerably higher.
Turning to the yields obtained with Eq.~\ref{EQ:fitBoltzAni}, the result for K$^-$
agrees well with that presented in~\cite{Pias15} and obtained with a different model.
The novelty is the yield for K$^+$, obtained for the first time 
for this colliding system and centrality.
Due to the similarity of $\langle A_{\text{part}} \rangle_{\text{b}}$ to the data 
from Al+Al collisions at the same beam energy 
($46.5 \pm 2.0$ for Ni+Ni vs 42.5 for Al+Al~\cite{Gasi16,Pias16}),
these data sets may be juxtaposed. Despite somewhat different acceptances and spectator sizes,
all the parameters obtained in our procedure are in good agreement with those
shown in Tab.~2 of Ref.~\cite{Gasi16}, if the systematic errors are included.

The ratio of kaon yields was extracted from the above fit results, 
and was found to be:

\begin{equation}
\frac{\text{P(K}^-)}{\text{P(K}^+)} ~=~
\left( 2.53 \pm 0.06 \pm 0.06 \right) \times 10^{-2}
\end{equation}

\noindent This value is in good agreement with the result obtained 
by the KaoS collaboration~\cite{Menz00}, 
measured within a narrower and more forward-shifted acceptance.
Our result is obtained within a wider acceptance and at higher precision.

\begin{table*}[tb]
 \caption{Parameters of the best fit of Eq.~\ref{EQ:fitBoltzAni} 
 to the phase space distribution of charged kaons emitted from Ni+Ni collisions at 1.91A GeV.
 The first uncertainty is statistical, while the second is systematic.
 See text for details.}
 \label{Tab:KaonFit}
 \begin{tabular}{ccccccc}
  \hline
  Particle &        $N$ (mul. per triggered event)  & &     $T_{\text{eff}}$   [MeV]                & &     $a_2$                   & $\chi^2   \slash \nu$   \\
  \hline
  K$^+$    & \quad $(3.598 \pm 0.012 \pm 0.043) \times 10^{-2}$ & & \quad $   110.9 \pm 0.6 \pm  0.4$ \quad & & $0.430 \pm 0.016 \pm 0.013$ &      36.2                 \\
  K$^-$    & \quad $(9.1   \pm 0.2   \pm 0.2  ) \times 10^{-4}$ & & \quad $\ \ 71.3 \pm 2.6 \pm  9.0$ \quad & & $0.16\ \pm 0.08\ \pm 0.11\  $ &       2.5                               \\
  \hline
 \end{tabular}
\end{table*}

\subsection{Subtraction of the $\phi$ meson contribution from the K$^-$ spectra}

For the analysed experiment $\phi$ mesons were found to be produced 
with a yield comparable to that of K$^-$:
\mbox{$\text{P}(\phi) \slash \text{P(K}^- )$} $ = 0.44 \pm 0.07 ^{+0.16}_{-0.10}$~\cite{Pias15}. 
Assuming that the vacuum value of 
$\text{BR} (\phi \rightarrow \text{K}^+\text{K}^-) = 48.9\%$~\cite{PDG}
remains the same for $\phi$ mesons produced in the collision zone of heavy ions, 
this means that about 22\% of negative kaons originate from decays of $\phi$. 
The kinematic properties of these K$^-$ mesons are different to these
for K$^-$s emitted directly from the collision zone.
Also, some transport models aiming at extraction of kaon in-medium effects 
may not reproduce the $\phi$ meson contribution well.
Therefore, it might be of interest to obtain the ratio of 
charged kaon yields unaffected by the contribution from $\phi$ mesons. 
This procedure was performed as for the case of Al+Al collisions at 1.9A~GeV,
described in Sect.~6.2 of Ref.~\cite{Gasi16}. 
In brief, the $\phi$ mesons were sampled within the PLUTO package~\cite{PLUTO}
from the isotropic thermal distribution characterized by a temperature 
of $106 \pm 18 ^{+18}_{-14}$~MeV, as reported in~\cite{Pias15}.
They decayed into K$^+$K$^-$ pairs,
from which the phase space distribution of negative kaons was obtained.
This was subsequently subtracted from the experimental spectrum, 
shown in Fig.~\ref{fig:EkinReco}, with values of BR and 
\mbox{$\text{P}(\phi) \slash \text{P(K}^- )$} as described earlier.
The resulting distribution of the ratio of yields of charged kaons without
the $\phi$ meson contribution is presented in the right-hand panels 
of Fig.~\ref{fig:kmkpratio}.
Within the subtraction procedure the statistical and systematic uncertainties 
of the $\phi$ meson yield and temperature were accounted for.
In order to check whether the obtained distribution of the ratio of yields
exhibits some drop with kinetic energy it was fitted with:
(a) constant and 
(b) linear functions of this quantity.
The fit was applied globally to all the points.
For scenario (a) \mbox{$\chi^2 \slash \nu$} was found to be $4.2 \pm 1.5$~(syst).
For the linear approach (b) the slope was found to be $-0.106 \pm 0.014 \pm 0.021$, 
at \mbox{$\chi^2 \slash \nu$} $= 1.4 \pm 0.3$~(syst).
The contributions to the systematic errors comprised all the factors
specifed in case of the evaluation of the K$^- \slash \text{K}^+$ ratio, 
and the systematic uncertainties associated with the
yield and inverse slope of the $\phi$ meson emission. 
Although the average value of $\chi^2 \slash \nu$ is better for the linear scenario,
the assumption of a constant ratio cannot be rejected based on the current data.

\subsection{$\Lambda(1520)$ contribution}

Recently, it has been conjectured ~\cite{Lore17} that the relevant contribution 
to the K$^-$ spectrum at SIS18 energies might be the
$\Lambda(1520) \rightarrow \text{pK}^-$ decays (BR~=~22.5\%~\cite{PDG}). 
A contribution from this channel has not yet been analysed by FOPI,
but it can be estimated with the help of the Statistical Hadronization Model 
(SHM)~\cite{Andr10,Beca01}.
Six ratios of particle production yields have been selected
(c.f. Table~\ref{Tab:YieldRatios})
taking care that they were obtained at the same centrality
as the data presented in this paper. 
A least-squares fit of the SHM to these ratios and the 
$\langle A_{\text{part}} \rangle_{\text{b}}$ value
was performed using the THERMUS code~\cite{THERMUS}.
Particles with non-zero strangess were treated within the canonical ensemble, 
and the grand canonical approach was applied to the other hadrons. 
To stabilize the procedure the volume was fixed by an arbitrary value of the radius
(3~fm), whereas the canonical radius $R_\text{C}$ was subject to free fitting.
The algorithm resulted in the following values of the parameters: 
temperature $T = 76.1 \pm 0.5$~MeV, 
baryo-chemical potential $\mu_\text{B} = 821.5 \pm 1.8$~MeV, 
and $R_\text{c} = 2.10 \pm 0.05$~fm, 
found at \mbox{$\chi^2\slash\nu = 1.8$}.
As the procedure aimed only at estimating the $\Lambda(1520)$ yield, 
the systematic errors were not investigated. 
Our result is compatible at the 1$\sigma$ level with the values obtained for
Ar+KCl at 1.76A GeV~\cite{Agak11}. However, both these results
are slightly above the general trend of the data observed in Fig.~7 
of Ref.~\cite{Agak11}.

Within this calculation the ratio of $\Lambda(1520) \slash \text{K}^-$ yields
was found to be 0.46, which translates into a 10\% contribution 
of $\Lambda(1520)$ decays to the spectrum of negative kaons.
Comparing with the 22\% contribution from $\phi$ meson decays, 
one may conclude that the $\Lambda(1520) \rightarrow p\text{K}^-$ channel 
may have moderate relevance. 
An experimental investigation of this channel is advisable.

\begin{table}[tb]
 \caption{Ratios of yields of particles emitted from Ni+Ni collisions at 
  1.91A GeV analysed in this paper.
  The average number of participating nucleons was estimated 
  within the geometrical model to be 
  $\langle A_{\text{part}} \rangle_{\text{b}} = 46.5 \pm 2.0$.}

 \label{Tab:YieldRatios}
 \begin{tabular}{ccccc}
  \hline
  Quantity                 & $\quad$ &  Value                           & $\quad$ & Reference                \\
  \hline
  K$^- \slash \text{K}^+$  &         & $(2.53 \pm 0.08) \times 10^{-2}$ &         & This work                \\
  K$^+ \slash \pi^+$       &         & $(7.59 \pm 0.49) \times 10^{-3}$ &         & This work, \cite{Pelt97} \\
  $\pi^+ \slash \pi^-$     &         & 1.00 $\pm$ 0.08                  &         & \cite{Pelt97}            \\
  K$^0 \slash \Lambda$     &         & 0.78 $\pm$ 0.18                  &         & \cite{Mers07}            \\
  $\Lambda \slash \pi^-$   &         & $(1.09 \pm 0.13) \times 10^{-2}$ &         & \cite{Mers07,Pelt97}     \\
  $\phi \slash \text{K}^-$ &         & 0.44 $\pm$ 0.15                  &         & \cite{Pias15}            \\
  \hline
 \end{tabular}
\end{table}

\section{Summary}

We have presented the phase space distributions of K$^+$ and K$^-$ mesons 
as well as of the K$^- \slash \text{K}^+$ ratio 
from the central and semi-central collisions of Ni+Ni at the a energy of 1.91A GeV, 
measured within wide acceptance by the FOPI apparatus. 
An overall value of this ratio was found to be:
$\left( 2.53 \pm 0.06 \pm 0.06 \right) \times 10^{-2}$.

The data are tabulated for convenient comparison with the 
predictions of transport models, with the hope of a more precise extraction 
of the parameters quantifying the in-medium modification 
of the properties of charged kaons.

Benefitting from the $\phi$ meson data measured in the same experiment,
we also present the distribution of the \mbox{K$^- \slash \text{K}^+$} ratio 
obtained after subtraction of the contribution of 
$\phi \rightarrow \text{K}^+\text{K}^-$ decays to the K$^-$ spectra.

The corrected K$^- \slash \text{K}^+$ ratio seems to decrease with kinetic energy, 
confirmed by the low $\chi^2 \slash \nu$ value of the fit. 
However, due to the precision of the results obtained, the hypothesis of a 
constant ratio cannot be rejected. Using the measured particle ratios, 
we applied the Statistic Hadronization Model to estimate the contribution of 
$\Lambda(1520)$ decays to the K$^-$ spectra. It was found to be 10\%, 
which suggests that an experimental investigation of $\Lambda(1520)$
production may be relevant for the subsequent reduction of anti-kaon 
yields not originating from the hot and dense collision zone.

In addition, an analysis of the phase space distributions of K$^\pm$ mesons
provided the multiplicities, inverse slopes and polar anisotropy coefficients.
The inverse slope for K$^-$ was found to be clearly lower than that for K$^+$.
With the exception of the $a_2$ coefficient for K$^+$, 
the presented results are in line with previously published data.

\begin{acknowledgments}
This work was supported by the German BMBF Contract No. 05P12VHFC7, 
the National Research Foundation of Korea (NRF) under grant No. 2018R1A5A1025563,
the German BMBF Contract No. 05P12RFFCQ, 
the Polish Ministry of Science and Higher Education (DFG/34/2007), 
the agreement between GSI and IN2P3/CEA, 
the HIC for FAIR, 
the Hungarian OTKA Grant No. 71989, 
NSFC (Project No. 11079025), 
DAAD (PPP D/03/44611), 
DFG (Project 446-KOR-113/76/04) and 
the EU, 7th Framework Program, Integrated Infrastructure: 
  Strongly Interacting Matter (Hadron Physics), Contract No. RII3-CT-2004-506078.
\end{acknowledgments}

\appendix
\section{Experimental data points}
\label{Sect:Data}

\begin{table*}[ht]
 \caption{Measured values of the phase space distribution of charged kaons and the
  P(K$^-$) $\slash$ P(K$^+$) ratio as a function of kinetic energy and
  $\cos \vartheta_{\text{NN}}$ in the NN frame. 
  The first uncertainty is statistical, the second systematic.
  See text for details.
 }
 \label{Tab:KaonYields}
  \begin{tabular}{ccccc}
   \hline
   $E^{\text{kin}}_{\text{NN}} ~\text{[GeV]}$    &
   P(K$^+$) $\times~ 10^4$  &
   P(K$^-$) $\times~ 10^6$  &
   \quad P(K$^-$) $\slash$ P(K$^+$) $\times~ 10^2$ &
   \quad P(K$^-_{\text{direct}}$) $\slash$ P(K$^+$) $\times~ 10^2$ \\
   \hline
 & & -0.95 $ < \vartheta_{\text{NN}} < $ -0.8 & \\
   \hline
0.07 &                                 &   9.71 $ \pm $ 1.60 $ \pm $ 1.01  &                                   &                                  \\
0.10 &                                 &   8.76 $ \pm $ 0.85 $ \pm $ 0.62  &                                   &                                  \\
0.13 &                                 &   6.17 $ \pm $ 0.79 $ \pm $ 0.45  &                                   &                                  \\
0.16 & 1.85 $ \pm $ 0.04 $ \pm $ 0.03  &   4.92 $ \pm $ 0.54 $ \pm $ 0.40  &   2.67 $ \pm $ 0.30 $ \pm $ 0.22  &  2.00 $ \pm $ 0.32 $ \pm $ 0.28  \\
0.19 & 2.22 $ \pm $ 0.04 $ \pm $ 0.03  &   4.63 $ \pm $ 0.51 $ \pm $ 0.19  &   2.08 $ \pm $ 0.23 $ \pm $ 0.10  &  1.69 $ \pm $ 0.25 $ \pm $ 0.14  \\
0.22 & 2.17 $ \pm $ 0.04 $ \pm $ 0.04  &   3.69 $ \pm $ 0.57 $ \pm $ 0.42  &   1.70 $ \pm $ 0.27 $ \pm $ 0.20  &  1.41 $ \pm $ 0.28 $ \pm $ 0.21  \\
0.25 & 2.04 $ \pm $ 0.04 $ \pm $ 0.02  &   3.64 $ \pm $ 0.76 $ \pm $ 0.63  &   1.79 $ \pm $ 0.37 $ \pm $ 0.31  &  1.57 $ \pm $ 0.33 $ \pm $ 0.31  \\
0.28 & 2.00 $ \pm $ 0.04 $ \pm $ 0.01  &   3.36 $ \pm $ 0.68 $ \pm $ 0.66  &   1.68 $ \pm $ 0.34 $ \pm $ 0.32  &  1.53 $ \pm $ 0.34 $ \pm $ 0.33  \\
   \hline
 & & -0.80 $ < \vartheta_{\text{NN}} < $ -0.65 & \\
   \hline
0.07 &                                 &  11.84 $ \pm $ 0.90 $ \pm $ 0.44  &                                   &                                  \\
0.10 & 2.91 $ \pm $ 0.04 $ \pm $ 0.06  &   9.48 $ \pm $ 0.65 $ \pm $ 0.20  &   3.26 $ \pm $ 0.23 $ \pm $ 0.10  &  2.46 $ \pm $ 0.26 $ \pm $ 0.21  \\
0.13 & 2.62 $ \pm $ 0.03 $ \pm $ 0.05  &   8.78 $ \pm $ 0.60 $ \pm $ 0.14  &   3.35 $ \pm $ 0.23 $ \pm $ 0.08  &  2.70 $ \pm $ 0.25 $ \pm $ 0.17  \\
0.16 &                                 &                                   &                                   &                                  \\ 
0.19 & 2.04 $ \pm $ 0.03 $ \pm $ 0.03  &   4.66 $ \pm $ 0.90 $ \pm $ 1.65  &                                   &                                  \\
0.22 & 1.98 $ \pm $ 0.03 $ \pm $ 0.03  &   4.83 $ \pm $ 1.00 $ \pm $ 2.28  &                                   &                                  \\
0.25 & 1.78 $ \pm $ 0.04 $ \pm $ 0.03  &   3.18 $ \pm $ 0.95 $ \pm $ 1.83  &                                   &                                  \\
0.28 & 1.59 $ \pm $ 0.04 $ \pm $ 0.04  &                                   &                                   &                                  \\
   \hline
 & & -0.65 $ < \vartheta_{\text{NN}} < $ -0.50 \\
   \hline
0.07 & 2.79 $ \pm $ 0.04 $ \pm $ 0.06  &  13.20 $ \pm $ 0.91 $ \pm $ 0.34  &   4.74 $ \pm $ 0.33 $ \pm $ 0.15  &  3.67 $ \pm $ 0.37 $ \pm $ 0.30  \\
0.10 & 2.66 $ \pm $ 0.03 $ \pm $ 0.05  &   9.42 $ \pm $ 0.62 $ \pm $ 0.13  &   3.54 $ \pm $ 0.24 $ \pm $ 0.06  &  2.67 $ \pm $ 0.27 $ \pm $ 0.22  \\
0.13 & 2.51 $ \pm $ 0.03 $ \pm $ 0.04  &   8.01 $ \pm $ 0.57 $ \pm $ 0.21  &   3.19 $ \pm $ 0.23 $ \pm $ 0.06  &  2.49 $ \pm $ 0.26 $ \pm $ 0.18  \\ 
0.16 & 2.21 $ \pm $ 0.03 $ \pm $ 0.04  &   6.97 $ \pm $ 0.59 $ \pm $ 0.27  &   3.15 $ \pm $ 0.27 $ \pm $ 0.14  &  2.58 $ \pm $ 0.29 $ \pm $ 0.20  \\
0.19 &                                 &                                   &                                   &                                  \\
0.22 &                                 &                                   &                                   &                                  \\
0.25 & 1.11 $ \pm $ 0.04 $ \pm $ 0.03  &   4.95 $ \pm $ 1.74 $ \pm $ 2.95  &                                   &                                  \\
0.28 & 0.81 $ \pm $ 0.03 $ \pm $ 0.02  &                                   &                                   &                                  \\
   \hline
 & & -0.50 $ < \vartheta_{\text{NN}} < $ -0.35 & \\
   \hline
0.07 & 2.73 $ \pm $ 0.04 $ \pm $ 0.06  &   9.45 $ \pm $ 1.00 $ \pm $ 0.22  &   3.46 $ \pm $ 0.37 $ \pm $ 0.13  &  2.38 $ \pm $ 0.41 $ \pm $ 0.29  \\
0.10 & 2.52 $ \pm $ 0.03 $ \pm $ 0.01  &   8.69 $ \pm $ 0.63 $ \pm $ 0.27  &   3.45 $ \pm $ 0.25 $ \pm $ 0.11  &  2.54 $ \pm $ 0.29 $ \pm $ 0.24  \\
0.13 & 2.30 $ \pm $ 0.03 $ \pm $ 0.03  &   6.88 $ \pm $ 0.57 $ \pm $ 0.18  &   2.99 $ \pm $ 0.25 $ \pm $ 0.08  &  2.25 $ \pm $ 0.28 $ \pm $ 0.20  \\
0.16 & 2.10 $ \pm $ 0.03 $ \pm $ 0.02  &   5.56 $ \pm $ 0.52 $ \pm $ 0.27  &   2.65 $ \pm $ 0.25 $ \pm $ 0.13  &  2.05 $ \pm $ 0.28 $ \pm $ 0.20  \\
0.19 & 1.80 $ \pm $ 0.03 $ \pm $ 0.02  &   3.88 $ \pm $ 0.49 $ \pm $ 0.33  &   2.16 $ \pm $ 0.27 $ \pm $ 0.18  &  1.66 $ \pm $ 0.30 $ \pm $ 0.22  \\
0.22 & 1.56 $ \pm $ 0.02 $ \pm $ 0.02  &                                   &                                   &                                  \\
0.25 & 1.33 $ \pm $ 0.03 $ \pm $ 0.02  &                                   &                                   &                                  \\
   \hline
 & & -0.35 $ < \vartheta_{\text{NN}} < $ -0.20 & \\
   \hline
0.07 &                                 &                                   &                                   &                                  \\
0.10 & 2.45 $ \pm $ 0.03 $ \pm $ 0.02  &  10.09 $ \pm $ 0.81 $ \pm $ 0.50  &   4.12 $ \pm $ 0.33 $ \pm $ 0.19  &  3.18 $ \pm $ 0.36 $ \pm $ 0.29  \\
0.13 & 2.31 $ \pm $ 0.03 $ \pm $ 0.02  &   7.09 $ \pm $ 0.58 $ \pm $ 0.23  &   3.07 $ \pm $ 0.26 $ \pm $ 0.10  &  2.32 $ \pm $ 0.28 $ \pm $ 0.21  \\
0.16 & 2.12 $ \pm $ 0.03 $ \pm $ 0.02  &   5.88 $ \pm $ 0.62 $ \pm $ 0.34  &   2.78 $ \pm $ 0.30 $ \pm $ 0.16  &  2.18 $ \pm $ 0.32 $ \pm $ 0.22  \\
0.19 & 1.85 $ \pm $ 0.03 $ \pm $ 0.02  &   3.62 $ \pm $ 0.56 $ \pm $ 0.59  &   1.96 $ \pm $ 0.30 $ \pm $ 0.31  &  1.47 $ \pm $ 0.33 $ \pm $ 0.34  \\
0.22 & 1.68 $ \pm $ 0.03 $ \pm $ 0.02  &   2.41 $ \pm $ 0.48 $ \pm $ 0.48  &   1.43 $ \pm $ 0.28 $ \pm $ 0.28  &  1.05 $ \pm $ 0.31 $ \pm $ 0.29  \\
0.25 & 1.34 $ \pm $ 0.03 $ \pm $ 0.01  &   2.01 $ \pm $ 0.43 $ \pm $ 0.53  &   1.50 $ \pm $ 0.32 $ \pm $ 0.39  &  1.17 $ \pm $ 0.34 $ \pm $ 0.40  \\
   \hline
 & & -0.20 $ < \vartheta_{\text{NN}} < $ -0.05 & \\
   \hline
0.07 &                                 &                                   &                                   &                                  \\
0.10 & 2.69 $ \pm $ 0.04 $ \pm $ 0.03  &   9.96 $ \pm $ 1.56 $ \pm $ 0.61  &   3.71 $ \pm $ 0.58 $ \pm $ 0.24  &  2.85 $ \pm $ 0.60 $ \pm $ 0.31  \\
0.13 & 2.42 $ \pm $ 0.03 $ \pm $ 0.02  &   6.10 $ \pm $ 0.68 $ \pm $ 0.40  &   2.53 $ \pm $ 0.28 $ \pm $ 0.16  &  1.81 $ \pm $ 0.30 $ \pm $ 0.23  \\
0.16 & 2.30 $ \pm $ 0.03 $ \pm $ 0.02  &   4.14 $ \pm $ 0.58 $ \pm $ 0.46  &   1.80 $ \pm $ 0.25 $ \pm $ 0.20  &  1.26 $ \pm $ 0.28 $ \pm $ 0.24  \\
0.19 & 1.93 $ \pm $ 0.03 $ \pm $ 0.02  &   3.69 $ \pm $ 0.55 $ \pm $ 0.41  &   1.92 $ \pm $ 0.29 $ \pm $ 0.21  &  1.45 $ \pm $ 0.30 $ \pm $ 0.24  \\
0.22 & 1.50 $ \pm $ 0.03 $ \pm $ 0.01  &   2.55 $ \pm $ 0.70 $ \pm $ 0.86  &   1.70 $ \pm $ 0.47 $ \pm $ 0.57  &  1.28 $ \pm $ 0.49 $ \pm $ 0.58  \\ 
0.25 & 1.14 $ \pm $ 0.02 $ \pm $ 0.01  &                                   &                                   &                                  \\
   \hline
  \end{tabular}
\end{table*}

\end{document}